\documentstyle[aps]{revtex}
\begin{document}
\draft
\preprint{RU9241}
\title
{Orbital M1 versus E2 strength in deformed nuclei:\\
A new energy weighted sum rule}

\author{E. Moya de Guerra}

\address
{Instituto de Estructura de la Materia,\\
Consejo Superior de Investigaciones Cient\'{\i}ficas,\\
Serrano 119--123, 28006--MADRID, SPAIN}

\author{L. Zamick}

\address
{Department of Physics and Astronomy,\\
Rutgers University, Piscataway,\\
New Jersey 08855, USA}

\date{ }

\maketitle

\begin{abstract}
Within the unified model of Bohr and Mottelson we derive the
following linear energy weighted sum rule for low energy orbital
$1^+$ excitations in even--even deformed nuclei

\[
S^{lew}_{LE} (M_1^{orb}) \cong (6/5) \epsilon (B(E2; 0^+_1
\rightarrow 2_1^+ K=0)/Z e^2<r^2>^2) \mu^2_N
\]

\noindent
with $B(E2)$ the $E2$ strength for the transition from the ground
state to the first excited state in the ground state rotational
band, $<r^2>$ the charge r.m.s. radius squared and $\epsilon$ the
binding energy per nucleon in the nuclear ground state. It is
shown that this energy weighted sum rule is in good agreement
with available experimental data. The sum rule is derived using
a simple ansatz for the intrinsic ground state wave function
that predicts also high energy $1^+$ strength at $2 \hbar w$
carrying 50\% of the total $m_1$ moment of the orbital $M1$ operator.
\end{abstract}
\pacs{PACs Number(s): ?????}

\section{Introduction}

Over the last years much experimental and theoretical work has
been devoted to the study of orbital $1^+$ excitations~\cite{1}.
Thanks to the combined analyses of ($e,e'$), ($\gamma, \gamma'$)
and ($p,p'$) experiments the occurrence of low lying orbital
$1^+$ excitations in even--even deformed nuclei is now a well
documented and established fact. Furthermore it has been
recently found that the summed strength of the observed
excitations, that typically appear concentrated around an
average excitation energy $\sim$ 3 $MeV$, varies quadratically
with quadruple deformation~\cite{2}. This last observation has motivated
theoretical work~\cite{3,4,5,6,7,8} to understand the origin of the
quantitative relation between the orbital $M1$ strength and the
quadruple deformation parameter $\beta$ from microscopic
calculations~\cite{3,4,5} and to derive general formulas connecting
$M1$ and $E2$ strengths from simplified models~\cite{6,7,8}.

In Ref. [3] results were presented for summed orbital 1$^+$
strengths in $Sm$ and $Nd$ isotopes that had been obtained from
deformed $HF+BCS$ calculations with $SK III$ interaction,
using the angular momentum projection formalism and substracting
spurious contributions~\cite{9,10}. The results obtained including all
possible 2 quasi--particle
excitations up to $E_x=4 MeV$ were found to be proportional to
$\beta^2$, to agree with experimental data, and to account for
about 50\% of the total summed strength (obtained including all
possible 2 quasi--particle excitations up to $E_x \sim 25
MeV)$~\cite{3}. It was also found that for the deformed isotopes
the stronger contributions occur in the range $2MeV \leq E_x
\leq 4 MeV$ and that at higher excitations the strength is much
more fragmented. Because calculations neglecting pairing lead to
summed strengths proportional to $\beta$ rather than to
$\beta^2$ it was argued that dependence on $\beta^2$
is due to the combined effect of pairing and
deformation~\cite{3}. Similar results and conclusions have been
reached from QRPA calculations based on deformed Woods--Saxon
potentials with pairing, quadrupole--quadrupole and spin--spin
residual interactions~\cite{4,5}. The main difference is that the
above mentioned QRPA calculations find less orbital 1$^+$
strengths in the high energy region. We shall come back to this
point latter on.

In an attempt to establish a general connection between orbital
$M1$ and $E2$ strengths, linear energy weighted sum rules $(L E
W S R)$ for orbital 1$^+$ excitations have been recently
discussed by Heyde and De Coester~\cite{6} and by Zamick and
Zheng~\cite{7}. Within the framework of the interacting boson
model (IBM--2) Heyde and De Coester obtain

\begin{eqnarray}
\sum_f B(M1; 0^+_1 \rightarrow 1^+_f) E_x(1^+_f) = \sum_f B(E2;
0^+_1 \rightarrow 2^+_f)
\label{1}
\end{eqnarray}

\noindent
with an effective $E2$ charge appropriate for Boson models.

On the other hand Zamick and Zheng, using a
quadrupole--quadrupole interaction, find that the LEWSR
for orbital $M1$ is proportional to
the \underline{difference} of summed isoscalar and
summed isovector $E2$ strengths. In this work we
shall show that this approach can be carried further. This will
be done by introducing the mean field into the picture.
We will for the first time look at both the low energy $(\Delta
N=0)$ and the high energy $(\Delta N=2)$ contributions to LEWSR.

We shall here consider two closely related models for
calculating LEWSR. First a two body quadrupole--quadrupole
interaction is used to evaluate the double commutator

\begin{eqnarray}
\begin{array}{ll}
S^{lew} (M1^{orb}) & =\frac{3}{4 \pi} \sum_f (E_f-E_{g.s})|<f|
\vec{\mu}^{orb} |g.s>|^2 \mu^2_N\\
\\
 & = \frac{3}{8 \pi} \mu^2_N<g.s.|
\left[\vec{\mu}^{orb},\left[H,\vec{\mu}^{orb}\right] \right] |g.s.>
\end{array}
\label{2}
\end{eqnarray}

We note that the commutator in Eq. (2) vanishes for a pure
pairing interaction between like nucleons (i.e. between protons
only and neutrons only) and for a pure spin--spin interaction.
Therefore the LEWSR, Eq. (2), should not change when said
interactions are explicitly considered.
The strength of the quadrupole--quadrupole interaction is obtained from a self
consistency condition. In the second method one evaluates this
same double commutator using the one body deformed field.

\section{Linear Energy Weighted Sum Rule with a
Quadrupole--Quadrupole Interaction}

We write the quadrupole--quadrupole interaction as

\begin{eqnarray}
H_Q=-\chi \sqrt{5} \sum_{i>j} [(r^2 Y^2)_i (r^2 Y^2)_j]^0
\label{3}
\end{eqnarray}

As shown by Zheng and Zamick~\cite{7}, the value of LEWSR with this
interaction is (see also Ref. [16])

\begin{eqnarray}
\begin{array}{ll}
S^{lew} (M1) &\\
\\
 =\frac{9}{16\pi} \chi [\sum B(E2;1,1)-\sum B(E2;1,-1)]&
\end{array}
\label{4}
\end{eqnarray}

\noindent
where $B(E2; e_p, e_n)$ is the value of $B(E2)$ for the hypothetical operator

$$ e_p \sum_{\pi} r^2 Y^2 + e_n \sum_{\nu} r^2 Y^2 $$

The transition is from the $J=0^+$ ground state to excited
$J=2^+$ states. Eq. (4) holds if we add to the Q.Q interaction a
pairing interaction between like nucleons.

For classification purposes it should be noted that for an $N=Z$
nucleus with a $T=0$ ground state the
value of $B(E2,1,1)$ is four times that of the true electric
operator $B(E2,1,0)$ for $T=0$ final states and is zero for
$T=1$ final states. On the other hand $B(E2,1,-1)$ will reach
only $T=1$ final states and is four times the value of $B(E2)$
for the electric probe with $e_p=1$ and $e_n=0$.

A renormalized value of $\chi$, which we call $\chi_R$ can be
determined,
as discussed by Bes and Sorensen~\cite{14}. The
interaction of a single valence particle with all the other
particles, assuming they have an axially symmetric density
distribution with the $Z$ axis as a symmetry axis is

\begin{eqnarray}
H_{DF}=-\chi_R \frac{5}{16 \pi} Q_M(core) (2z^2-x^2-y^2)
\label{5}
\end{eqnarray}

\noindent
where $Q_M (core)$ is the mass Quadrupole moment of the core. It
is further shown that the quadrupole moment of the core is equal
to that of the valence nucleons. Thus the intrinsic mass
quadrupole moment $Q_0^M$ is equal to

\begin{eqnarray}
Q_0^M=Q^M(core) + Q(valence) \cong 2 Q^M(core)
\label{6}
\end{eqnarray}

If in contrast to the above we use a deformed oscillator
hamiltonian

\begin{eqnarray}
H_{DO}=\Sigma\left[\frac{p^2}{2m} + \frac{1}{2}m w_x^2
(x^2+y^2)+\frac{1}{2} m w_z^2 z^2 \right],
\label{7}
\end{eqnarray}

\noindent
then we can make it look like the above expression by introducing
the oscillator deformation parameter $\delta_0$

\begin{eqnarray*}
\omega_x^2=\omega_0^2(\delta_0) \left[1+\frac{2}{3} \delta_0\right]
\end{eqnarray*}

\begin{eqnarray}
\omega_z^2=\omega_0^2(\delta_0) \left[1-\frac{4}{3} \delta_0\right]
\end{eqnarray}

\begin{eqnarray*}
H=\sum \left(\frac{p^2}{2m} + \frac{1}{2}m \omega_0^2
r^2\right) + H_{DF}\\  \nonumber
\end{eqnarray*}

\noindent
where the deformed field term is

\begin{eqnarray}
H_{DF}=-m \omega_0^2 \delta_0 (2z^2-x^2-y^2)/3
\label{9}
\end{eqnarray}

We define a deformation parameter $\delta$

\begin{eqnarray}
\delta=\frac{3}{4}\;\;\frac{Q_0^{\pi}}
{<R^2>_{\pi}}
\label{10}
\end{eqnarray}

\noindent
that can be experimentally determined from the measured charge
quadrupole moment and r.m.s. radius with bare nucleon charges,
and  assume equal deformation for neutrons $(\nu)$ and protons
$(\pi)$. By equating the two expressions for $H_{DF}$ we obtain the
following expression for $\chi_R$

\begin{eqnarray}
\chi_R = m \omega_0^2
\frac{\delta_0}{\delta}\;\;\frac{8 \pi}{5<R^2>_M} \cong \frac{m
\omega_0^2 4\pi}{5<R^2>_{\pi}} \delta_0/\delta
\label{11}
\end{eqnarray}

However, if we use the Q$\cdot$Q interaction amongst all the
nucleons in the nucleus, i.e. if we allow $\Delta$N=2 mixing, then the
value of $\chi$ to be used is $\chi_R$/2, as also discussed by Bes and
Sorensen~\cite{14} (see pages 143-144).

We further note that B(E2,1,1) to an isoscalar state is
\underline{four} times that for an electric probe ($e_p$ = 1,
$e_n$ = 0). Likewise with $e_p$ = 1, $e_n$ = $-$1
we reach the isovector states with four times the rate of what an
electric probe would give. Thus, in order to make easier comparisons
with electric probes we give the sum rule in the following way

\begin{eqnarray}
\begin{array}{ll}
S^{lew}(M1)_{\rm all~ particles} =  & \\
\\
 \frac{9}{10
}\frac{m\omega_o^2}{<R_{\pi}^2>}\frac{\delta_o}{\delta} \left[\frac{\sum
B(E2,1,1)}{4}- \frac{\sum B(E2,1,-1)}{4}\right].&
\end{array}
\label{12}
\end{eqnarray}

One final remark to this section. If we had taken for our
interaction $\frac{\chi}{2} \Sigma_{i,j} Q(i)\cdot Q(j)$, thus
keeping $i=j$ terms, we would reach the $SU(3)$ limit. The
$i=j$ terms add a single
particle term of the form $r^4$ to the potential and
cause a splitting of single particle energies with different $l$.
In the $SU(3)$ limit in a calculation involving only one
major shell (i.e. no $\Delta N=2$ mixing) the $E2$ operator
with $e_p=1, e_n=1$ reaches only one $2^+$ state--this can be
identified as the $J=2^+ K=0$ member of the ground state band.
The $E2$ operator with $e_p=1, e_n=-1$ (isovector) connects
$J=2^+ K=1$ member of the scissors
mode rotational band, as was first pointed out by Retamosa et
al.~\cite{11}. It should be noted that in shell model calculations the
scissors mode band can get fragmented into two bands, one with isospin
$T=T_0$ and one with isospin $T=T_0+1$, where $T_0$ is the g.s.
isospin. Thus the isovector $E2$ will be fragmented to two
states. Useful formulae for matrix elements of $Q.Q$ in the
Boson SU(3) scheme have been obtained by Rosenteel~\cite{12}

\section{The Double Commutator Method Applied Directly to the
deformed field}

We use a simple
approximation within the framework of the unified model of Bohr
and Mottelson~\cite{13} for even--even axially symmetric
rotational nuclei, where the properties of the ground state
rotational band are given in terms of properties of the
intrinsic ground state. For the intrinsic ground state we use the
anisotropic Harmonic Oscillator satisfying the selfconsistency
condition. The H.O. is used as an auxiliary model because ground
state expectation values have simple analytical expressions that
can then be reinterpreted in terms of those of the ``true''
intrinsic ground state, i.e., in terms of experimentally known
properties of the ground state band.

Hence in this section we consider a rotational model picture with
intrinsic hamiltonian of the form $H_{DO}$ in Eq. (7), subject
to the selfconsistency condition~\cite{13}

\begin{eqnarray}
\Sigma_x \omega_x=\Sigma_y \omega_y=\Sigma_z \omega_z=\Sigma \omega/3
\label{13}
\end{eqnarray}

In the rotational model picture the magnetic moment operator is
$\vec{\mu}=g_R \vec{I}+\vec{\mu}_{int}$, with $\vec{I}$ the total
angular momentum wihich does not contribute to magnetic
excitations and $g_R$ the collective gyromagnetic ratio that is
experimentally determined from the magnetic moment of $2^+,
4^+$ ... states in the ground state band. In even--even nuclei
$\vec{\mu}_{int}$ does not contribute to static magnetic moments
because it has zero spectation value in the intrinsic time even ground
state. The orbital part of $\vec{\mu}_{int}$ is

\begin{eqnarray}
\vec{\mu}_{int}^{orb}=\sum_i (g_l^i-g_R) \vec{l}_i=(1-g_R)
\vec{L}_{\pi} -g_R \vec{L}_{\nu}
\label{14}
\end{eqnarray}

Thus the evaluation of the double commutator in eq. (2) ammounts
to the evaluation of the double commutator of
$\vec{\mu}^{orb}_{int}$ with $H_{DO}$ in eqs. (14) and (7),
respectively.

Can we obtain a similar result to that obtained with a $Q \cdot Q$
interaction by applying the double commutator method
directly to the deformed one body field?

Using the fact that

\begin{eqnarray}
\begin{array}{ll}
\left[\vec{l}_i,
\left[\vec{l}_j,(2z^2-x^2-y^2)_k\right]\right]= & \\
\\
6(2z^2-x^2-y^2)_k \delta_{ik} \delta_{jk} &
\end{array}
\label{15}
\end{eqnarray}

\noindent
we readily find that

\begin{eqnarray}
\begin{array}{ll}
S^{lew}(M1^{orb}) = &\\
\\
 \frac{3\mu^2_N}{4\pi} \delta_0 m \omega^2_0
\left\{(1-g_R)^2 Q_0^{\pi}+g^2_R Q_0^{\nu}\right\}&
\end{array}
\label{16}
\end{eqnarray}

Using the definition of $\delta$ in eq. (10) and
the relationship for the $B(E2)$ for the ground
state $K=0$ band of an even--even nucleus

\begin{eqnarray}
B(E2; 0^+ \rightarrow 2^+ k=0) = \frac{5}{16 \pi} Q^2_{0 \pi}
\label{17}
\end{eqnarray}

We write eq. (16) in the form

\begin{eqnarray}
\begin{array}{ll}
S^{lew}(M1^{orb})= &\\
\\
 \frac{9 \mu^2_N\;m w_0^2}{10\;<R^2>_{\pi}}
\frac{\delta_0}{\delta} B(E2;0^+ \rightarrow 2^+ k=0) F_{\pi
\nu} &
\end{array}
\label{18}
\end{eqnarray}

\noindent
where

\begin{eqnarray}
F_{\pi \nu}=2\left[(1-g_R)^2+g_R^2 \frac{Q^{\nu}_0}{Q^{\pi}_0}\right]
\label{19}
\end{eqnarray}




We compare this result to that of the previous section for the
Q$\cdot$Q interaction. If we
remember that B(E2)$_{\pi} \approx $ B(E2,1,1)/4), the
differences between eqs. (18) and (12) are the presence of the
isovector $E2$ term in the latter and also the fact that eq. (18)
involves only the $B(E2)$ to the lowest $2^+$ state and contains
the factor $F_{\pi \nu}$.

We note that for a nucleus with equal proton and neutron bodies
$(g_R=1/2$, $Q_0^{\nu}/Q_0^{\pi}=1)$ the intrinsic orbital
operator in eq. (14) is purely isovector and the factor defined
in eq. (19) takes the value $1(F_{\pi \nu}=1)$, as is for
instance the case for $^{20}Ne$. In this case the dominant
contribution to eq. (12) is by far the $B(E2)$ value for the
transition to the lowest $2^+$ state, which in the $SU(3)$ limit
is the $J=2^+$ member of the ground state $(k=0)$ band and
exhaust the sum rule for quadrupole isoscalar transitions from
ground state.

Thus, in this particular case, one can see that the LEWSR in eq.
(18) is similar to that in eq. (12), but in the general case one
can only check by numerical comparison. One difference is of
course that eq. (18) applies to deformed nuclei with a well
developed rotational g.s. band, while eq. (12) is in principle
more general. We would like also to point out that the effective
nucleon gyromagnetic ratios $g^i_{eff}=(g^i_l-g_R)$ for dipole
magnetic excitations appearing in eq. (14) play a role analogous
to the effective charges for dipole electric excitations. In the
last case the effective charges result from the substraction of
the spourious center of mass motion, in the first case
$g^i_{eff}$ results from the substraction of the spourious rotation.
In what follows we shall see that within the
selfconsistent mean field picture, implementation of the
selfconsistency condition in eq. (13) allows to write eq. (18)
in a more practical way for phenomenological analysis.

\section{Additional Considerations}

First we note
that the results in Eqs. (16), (18) can also be obtained considering p.h.
contributions. But then we get additional insight. If we consider
p.h. contributions to the LEWSR

\begin{eqnarray}
\begin{array}{ll}
S^{lew}(M1^{orb})= &\\
\\
 \frac{3}{4\pi}\sum_{ph}(\epsilon_p-
\epsilon_h)|<ph|\vec{\mu}^{orb}_{int.}|0>|^2 \; \mu^2_N, &
\end{array}
\label{20}
\end{eqnarray}

\noindent
using the selection rules for $\vec{\mu}^{orb}_{int}$ in the H.O. basis
{}~\cite{16}, we get two sets of p.h. excitations. One set with
excitation energy $|\epsilon_p-\epsilon_h|$ $=|\hbar w_x-\hbar w_z|$
and one set with excitation energy $|\epsilon_p - \epsilon_h|$ $=
(\hbar w_x + \hbar w_z)$.

This allows to write Eq. (20) as

\begin{eqnarray}
S^{lew}(M1^{orb})=S_{LE}+S_{HE}
\label{21}
\end{eqnarray}

\noindent
with $S_{LE}$ the sum from all the low energy p.h. excitations
$(|\epsilon_p-\epsilon_h|=|\hbar w_x-\hbar w_z|$ $\simeq \hbar
w_0 \delta_0)$

\begin{eqnarray}
S_{LE}=\left[(1-g_R)^2 {\cal S}^{\pi}_{LE}+g^2_R {\cal S}^{\nu}_{LE}
\right]\mu^2_N
\label{22}
\end{eqnarray}

\begin{eqnarray}
{\cal S}_{LE}^{\rho}=\frac{3}{8 \pi} (\hbar w_x-\hbar
w_z)
\left(\Sigma_z^{\rho}-\frac{\Sigma^{\rho}_x+\Sigma^{\rho}_y}{2}\right)
(\beta^+)^2
\label{23}
\end{eqnarray}

\noindent
with $\rho=\pi, \nu$, and $S_{HE}$ the sum
from all the high energy p.h. excitations
$(|\epsilon_p-\epsilon_h|$ $=(\hbar w_x+\hbar w_z)$ $\simeq 2\hbar
w_0)$

\begin{eqnarray}
S_{HE}=\left[(1-g_R)^2 {\cal S}^{\pi}_{HE}+g_R^2 {\cal S}^{\nu}_{HE}
\right]\mu^2_N
\label{24}
\end{eqnarray}

\begin{eqnarray}
{\cal S}^{\rho}_{HE} =\frac{3}{8 \pi} (\hbar w_x+\hbar w_z)
\left(\Sigma^{\rho}_z+\frac{\Sigma^{\rho}_x+
\Sigma^{\rho}_y}{2}\right) (\beta^-)^2
\label{25}
\end{eqnarray}

\noindent
with

\[
\beta^{\pm} = \sqrt{\frac{w_x}{w_z}} \pm
\sqrt{\frac{w_z}{w_x}}
\]

Using the selfconsistency condition Eq. (13) with
$\Sigma=\Sigma^{\pi}$ $+\Sigma^{\nu}$ it
is a simple matter to show that

\[
{\cal S}^{\rho}_{LE}={\cal S}^{\rho}_{HE}=
\frac{\hbar \Sigma^{\rho} \omega}{8 \pi}\;\;
\frac{(\omega^2_x-\omega^2_z)^2}{\omega^2_x \omega^2_z}
\]

\begin{eqnarray}
=\frac{3}{8 \pi} \delta_0 m\; \omega_0^2 Q^{\rho}_0; \;\;\; \rho=\pi, \nu.
\label{26}
\end{eqnarray}

Where we have used the expresion for $Q^{\rho}_0$

\[
Q_0^{\rho}=\frac{\hbar}{m} \left(2
\frac{\Sigma^{\rho}_z}{\omega_z} -
\frac{\Sigma^{\rho}_x+\Sigma^{\rho}_y}{\omega_x}\right)
\]

\begin{eqnarray}
=\frac{2}{3}\;\; \frac{\hbar \Sigma^{\rho} \omega}{m} \;\;
\frac{(\omega^2_x-\omega^2_z)}{\omega^2_x \omega^2_z},
\label{27}
\end{eqnarray}

\noindent
and eq. (8) for $\omega_x, \omega_z$.

Therefore using eqs. (20) to (26) we find the interesting result
that

\begin{eqnarray}
S_{LE}=S_{HE}=\frac{1}{2} S^{lew}(M1^{orb})
\label{28}
\end{eqnarray}

\noindent
where $S^{lew} (M1^{orb})$ coincides with the value given in Eq. (16),
see also eq. (18).

Thus
within this approach the strength function has only two peaks
one at low energy $(E^{LE} \simeq \hbar w_0 \delta_0)$ and one
at high energy $(E^{HE} \simeq 2\hbar w_0)$. A pairing
interaction between like nucleons, and/or a spin--orbit
interaction will cause splitting in each of these peaks but will
hardly remove strength from the low energy peak to the high
energy peak and viceversa.
Therefore we write a LEWSR for the
low energy part of the strength function taking half the value
given in Eq. (16), or equivalently in eq. (18).

To write the final expression for the energy weighted sum rule
for low orbital $1^+$ excitations we substitute eq. (18) into
eq. (28). But first we replace the H.O. parameters $\omega_0, \delta_0$ as
well as $Q^{\nu}_0/Q^{\pi}_0$ in eqs. (18), (19) by
phenomenological parameters directly known from experimental
electromagnetic properties of the ground state band. To this end
we use additional relations that hold within the deformed H.O.
model satisfying the selfconsistency condition in eq. (13).

Contrary to $Q^{\pi}_0$, which is directly related to the
experimental quadrupole moments and $B(E2)$ values of the g.s.
band, there is no direct information on $Q^{\nu}_0$. Using the
definition of the collective gyromagnetic ratio $g_R={\cal
I}^{\pi}_{cr}/{\cal I}_{cr},$ with ${\cal I}_{cr}={\cal
I}^{\pi}_{cr}+{\cal I}^{\nu}_{cr}$ the cranking moment of
inertia, which in this simplified model is given by

\begin{eqnarray}
{\cal I}^{\rho}_{cr}=\frac{2}{3} \Sigma^{\rho} \omega
\frac{(\omega^2_x+\omega^2_z)}{\hbar \omega^2_x \omega^2_z},
\label{29}
\end{eqnarray}

\noindent
we can write

\begin{eqnarray}
g_R={\cal I}^{\pi}_{cr}/{\cal I}_{cr}=\Sigma^{\pi}/\Sigma
\label{30}
\end{eqnarray}

\noindent
and using eqs. (27) and (30) we find that

\begin{eqnarray}
\frac{Q^{\nu}_0}{Q^{\pi}_0}=(1-g_R)/g_R
\label{31}
\end{eqnarray}

On the other hand the charge r.m.s. radius squared in this model
is

\begin{eqnarray}
\begin{array}{ll}
<r^2>=\frac{<R^2>^{\pi}}{Z}=\frac{\hbar}{Z m}
\left(\frac{\Sigma_z}{\omega_z}+\frac{\Sigma_x+\Sigma_y}{\omega_x}
\right) & \\
\\
 =\frac{\hbar \Sigma^{\pi} \omega}{3mZ}
\left(\frac{\omega^2_x+2 \omega^2_z}{\omega^2_x
\omega^2_z}\right) &
\end{array}
\label{32}
\end{eqnarray}

\noindent
which together with eqs. (8), (10) and (30) give

\begin{eqnarray}
\delta_0=\delta/\left(1+\frac{2}{3}\delta\right)
\label{33}
\end{eqnarray}

\noindent
and

\begin{eqnarray}
m \omega^2_0 \frac{\delta_0}{\delta}= \frac{\hbar \Sigma^{\pi}
\omega}{Z <r^2>D (\delta)}=\frac{\hbar \Sigma
\omega}{AD(\delta)} \; \frac{A}{Z} \; \frac{g_R}{<r^2>}
\label{34}
\end{eqnarray}

\noindent
with

\[
D(\delta)=\left(1-\frac{2}{3} \delta\right) \left(1+\frac{4}{3}
\delta \right)
\]

\noindent
and $\delta$ fixed by the relations in eqs. (10) and (17), i.e.,

\begin{eqnarray}
\delta =\frac{3}{4} \sqrt{\frac{16{\pi}}{5}}
\left[\frac{B(E2;0^+_1 \rightarrow 2^+_1 k=0)}{e^2 Z^2
<r^2>^2}\right]^{\frac{1}{2}}.
\label{35}
\end{eqnarray}

We note that the experimental quadrupole deformation parameter
$\beta$ is usually defined in terms of $B(E2)$ and is related to
$\delta$ in Eq. (35) by $\beta=\sqrt{\frac{\pi}{5}} \frac{4}{3}
\delta$.

The selfconsistency condition Eq.(13), which ensures that the shape of the
potential follows the shape of the density, also ensures that i) the momentum
distribution is isotropic, and ii) the energy is minimum. These
intrinsic ground state properties are satisfied by the ground
state solution of deformed HF (or HF+BCS) calculations with
density dependent effective interactions. Thus $\Sigma^{\rho}_i,
w_i$, can be considered as
effective quantum numbers and H.O. frequencies
in the different directions corresponding to the
expectation values of $Q_o$ and $r^2$
in the ``true'' intrinsic ground state~\cite{15}. Similarly
$\left(\frac{3}{8} \hbar \Sigma \omega/A \right)$ can be
considered as the effective H.O. energy per
particle that corresponds to the true binding energy per
particle $(\epsilon =|Eg.s.|/A)$,

\begin{equation}
\hbar \Sigma \omega/A= 8 \epsilon/3
\label{36}
\end{equation}

Substituting Eqs. (31), (34), (36) into Eq. (18), and using Eq.
(28), we can finally write a LEWSR for the low energy
orbital $1^+$ excitations as

\begin{eqnarray}
S^{lew}_{LE} (M1^{orb})=G_{\pi \nu}\frac{6}{5}
\frac{Z \epsilon}{D(\delta)} \frac{B(E2; 0_1^+ \rightarrow 2^+_1k=0)}{(e
Z<r^2>)^2} \mu^2_N
\label{37}
\end{eqnarray}

\noindent
with

\begin{eqnarray}
G_{\pi \nu}=g_R (1-g_R) 2A/Z
\label{38}
\end{eqnarray}

For a nucleus with equal proton an neutron bodies $G_{\pi
\nu}=1$ because $g_R=1/2$ and $A/Z=2$. In practice most deformed
nuclei have $g_R<1/2, A/Z>2$ but $G_{\pi \nu}$ is still close to
1. For deformed rare earth nuclei since $g_R$ values are not
known with high accuracy one can use the approximation $G_{\pi
\nu} \simeq 1$ to evaluate the right hand side of eq. (37). This
approximation is used in the results shown in the last columm
of Table I. For each nucleus in the table typically the value of
$G_{\pi \nu}$ oscillates between 0.9 and 1.2, using $g_R$ values
compatible with the experimental data in Refs. [17] and [19].

In table I we show results for several deformed nuclei
obtained from Eq. (37) using
$B(E2)$ values from Ref. [17], $<r^2>$ values from Ref. [18] and
binding energies from Ref. [19] approximating $G_{\pi \nu}$ by one.
The results are compared to
available experimental data on several deformed nuclei. It is
interesting to see that this simple approximation leads to
results in good agreement with experimental data.
Therefore Eq. (37) can be
considered as a semiempirical LEWSR.

To the extent that more sophisticated microscopic calculations, as
those discussed in Refs. [3-5], are able to reproduce the low
energy part of the strength function for the orbital $M1$
operator, they also serve as a theoretical sound basis to support Eq. (37). As
mentioned in the introduction, the main difference between the
QRPA results of Refs. [4,5] and the results in Ref. [3], that
are obtained without inclusion of residual interaction, is that
the former find less orbital $1^+$ strength in the high energy
region, while the latter find $\sim 50\%$ of the scissors mode
strength in the energy interval $4 MeV \leq E_x \leq 25 MeV$
strongly fragmented in many 2 quasi--particle excitations. On the
contrary Zawischa and Speth~\cite{20}  performing QRPA
calculations with Migdal interaction find that the strongest
scissors like excitations take place at $E \simeq 22 MeV$.
Whether the ``true'' residual interaction tends to damp down the
orbital $1^+$ excitations in the high energy region, or to collect it back
into one or a few strong peaks,
is still an open question that cannot be answered
without experimental verification.
In this context it should be mentioned that the isovector term in Eq.
(12), which came from a $Q \cdot Q$ interaction, will provide some damping for
$\Delta N=2$ excitations~\cite{7}.
\section{Conclusions}

The main point of this work is to show that with a minimum of number
of assumptions one can obtain a simple expression (Eq. (37)) which
relates the energy weighted sum rule for orbital magnetic dipole
strength to the electric quadrupole strength, and hence to the
deformation parameter $\delta$. The simple expression involves well
known quantities - the binding energy per particle, the mean square
radius and the B(E2) and $g_R$. Although much work along these lines has been
done using the interacting boson approximation I.B.A. we find that we
can get equally simple expressions working directly with \underline{fermions}.
This indicates that the connection between the orbital M1 and B(E2) is
quite general and quite natural.

We further emphasize that the energy weighted orbital magnetic
strength should have a low energy part, to which our formula, as seen
in Table 1, gives very good fits and a high energy part ($\Delta N=2$).
In this rotational model approach the relative energy weighted
strengths are easy to calculate.

Our derivation using the one body field approach may not be as
rigorous as the derivation using the two--body interaction and
it is only applicable to deformed nuclei with a rotational
ground state band. The two body interaction
approach involving an interaction $\chi Q \cdot Q$ with $\chi$ chosen
selfconsistently gives a slightly different expression even for
$N=Z$ nuclei.
In the two body approach we get the difference between summed
isoscalar and isovector B(E2)'s. It should however be noted that with
a $Q \cdot Q$ interaction the isoscalar strength goes to only one
state in the $\Delta N=0$ SU (3) limit - the J=2$^+$ member of the K=0
ground state band. Thus if we drop the isovector term the two
approaches give the same answer. The isovector term in the $\Delta N=0$
SU (3) limit consists of an E2 transition to the K=1, J=2$^+$ member
of the scissors mode band. It would be of interest to try to measure
this B(E2) via electro excitation. However, calculations~\cite{7}
indicate that the isovector B(E2) is much weaker than the isoscalar
one and so indeed the main formula of this work (Eq. (37)) is quite
good.

In the future it will be of interest to extend these energy weighted
sum rule techniques to states of higher multipolarity.
It would also be of interest to extend the double commutator method
from schematic to realistic two body interactions.

\acknowledgments

We thank Dao Chen Zheng for useful comments. This work has been
supported in part by DGICYT (Spain) under contract PB87/0311.
One of us (L.Z.) acknowledges the support of the U.S. Dept. of
Energy, grant \# DE--FGO5--86ER--40299

\newpage
\mediumtext
\begin{table}
\caption{Comparison of experimental values for $\sum_x$
$E_x B(0^+ \rightarrow 1^+_x)$ with the results obtained from Eq.
(37) for the low energy linear energy weighted sum rule of
orbital $1^+$ strength (see text).
Also shown in the second column are
the experimental $B(E2)$ values from Ref. [17] used in this work.}
\begin{tabular}{cccccc}
{\rm Nucleus} &  B(E2) $\uparrow$ & $\displaystyle\sum_x E_x
B(0^+\rightarrow 1^+_x)(\mu^2_N MeV)$  & $S^{lew}_{LE} (M1^{orb})
(\mu^2_N MeV)$\\
& (be)$^2$ & Expt: $E_x \leq$ 4 MeV  & Theory\\
\tableline
$^{146}Nd$  & 0.76  & 2.005$^{\rm a}$   &  1.968 \\
$^{148}$Nd & 1.38  & 3.331$^{\rm a}$ &  3.339 \\
$^{150}$Nd & 2.75  & 5.95$^{\rm a,b}$  &6.439 \\
\\
$^{148}$Sm & 0.72  & 1.566$^{\rm c}$  &1.719 \\
$^{150}$Sm & 1.35  & 3.085$^{\rm c}$  &3.086 \\
$^{152}$Sm & 3.44  & 7.003$^{\rm c}$  &7.310 \\
$^{154}$Sm & 4.36  & 8.189$^{\rm c}$  &8.968 \\
\\
$^{156}$Gd & 4.64  & 8.039$^{\rm d}$  &8.946 \\
$^{158}$Gd & 5.02  & 8.287$^{\rm d}$  &9.616 \\
$^{160}$Gd & 5.25  & 7.246$^{\rm d}$  &9.819 \\
\\
$^{162}$Dy & 5.28  & 8.942$^{\rm b}$  &9.926 \\
$^{164}$Dy & 5.60  & 9.678$^{\rm e}$  &10.462
\end{tabular}
\tablenotetext[1]{Ref. [21].}
\tablenotetext[2]{Ref. [22].}
\tablenotetext[3]{Ref. [2].}
\tablenotetext[4]{Ref. [3].}
\tablenotetext[5]{Ref. [24]. The value quoted has been
obtained substructing the spin strength below $4\;\;MeV$ seen in
$(p,p')$ and reported in ref. [25].}
\end{table}
\end{document}